\ifdefined\pdfminorversion\pdfminorversion=7\fi
\ifdefined\pdfsuppresswarningdupdest\pdfsuppresswarningdupdest=1\fi
\documentclass[fleqn,usenatbib,useAMS,onecolumn]{mnras}

\usepackage[T1]{fontenc}

\IfFileExists{newtxtext.sty}{
  \IfFileExists{newtxmath.sty}{
    \usepackage{newtxtext,newtxmath}
  }{
    \IfFileExists{mathptmx.sty}{\usepackage{mathptmx}}{}
  }
}{
  \IfFileExists{mathptmx.sty}{\usepackage{mathptmx}}{}
}

\DeclareRobustCommand{\VAN}[3]{#2}
\let\VANthebibliography\thebibliography
\def\thebibliography{\DeclareRobustCommand{\VAN}[3]{##3}\VANthebibliography}

\IfFileExists{multirow.sty}{\usepackage{multirow}}{\providecommand{\multirow}[3]{#3}}
\IfFileExists{booktabs.sty}{\usepackage{booktabs}}{%
  \providecommand{\toprule}{\hline}
  \providecommand{\midrule}{\hline}
  \providecommand{\bottomrule}{\hline}
}
\IfFileExists{makecell.sty}{\usepackage{makecell}}{}
\usepackage{array}

\usepackage{graphicx}	% Including figure files
\usepackage{amsmath}	% Advanced maths commands
\usepackage{url}

\newcommand {\Ron}{\mathrm{Ro^*}}
\newcommand {\Dt}[1]{\frac{D #1}{D t}}
\newcommand {\bV}{\boldsymbol{v}}
\newcommand {\bOmega}{\boldsymbol{\Omega}}
\newcommand {\eR}{\mathbf{e_r}}
\newcommand {\ey}{\mathbf{e_\theta}}
\newcommand{\co}{\cos{\theta}}
\newcommand{\si}{\sin{\theta}}
\newcommand{\coe}[1]{\cos^{#1}{\theta}}
\newcommand{\sie}[1]{\sin^{#1}{\theta}}

\title[Slantwise Convection Scheme]{Parameterizing slantwise convection in icy moon oceans}

\author[Zeng and Jansen]{
Yaoxuan Zeng$^{1}$\thanks{E-mail: yxzeng@uchicago.edu}
and Malte F. Jansen,$^{1}$
\\
$^{1}$Department of the Geophysical Sciences, University of Chicago, Chicago, IL, 60637, USA
}

\date{Accepted XXX. Received YYY; in original form ZZZ}

\pubyear{\the\year{}}

\begin{document}
\label{firstpage}
\pagerange{\pageref{firstpage}--\pageref{lastpage}}
\maketitle
% \linenumbers

\begin{abstract}
Convection in icy moon oceans is strongly influenced by rotation, organizing into slantwise columnar structures aligned with the planetary rotation axis. They generate significant meridional heat transport, which can affect the ice shell topography, a primary observable of these moons. However, global ocean simulations cannot resolve convection under realistic icy moon conditions, and traditional convection schemes cannot represent slantwise convection. Here, we develop a slantwise convection scheme and implement it in a global ocean model. We perform benchmark tests in a global spherical shell by comparing parameterized fluxes with convection-resolving simulations. The scheme reproduces the meridional heat transport inside the tangent cylinder, where slantwise convection dominates. The resulting meridional heat transport significantly modifies the surface heat flux, producing variations comparable to the imposed bottom heating magnitude. Although the simulations with parameterized convection cannot fully reproduce the temperature structure, likely due to an inability to reproduce the temperature gradients near the boundaries, they capture the bulk interior vertical temperature gradient. The new scheme allows unresolved slantwise convection to be represented in global ocean simulations for icy moons. It is also applicable to other rapidly rotating oceans with small natural Rossby number ($\Ron \ll 1$), including deep ocean worlds on exoplanets.

\end{abstract}


%%%%%%%%%%%%%%%%%%%%%%%%%%%%%%%%%%%%%%%%%%%%%%%%%%

%%%%%%%%%%%%%%%%% BODY OF PAPER %%%%%%%%%%%%%%%%%%

\section{Introduction}\label{sec:intro}

In the past decades of outer solar system exploration, the Galileo and Cassini missions have revealed that several icy moons host global subsurface oceans \citep{khurana1998induced,neubauer1998sub,zimmer2000subsurface,kivelson2002permanent,iess2012tides,mckinnon2015effect,thomas2016enceladus}. On some moons, such as Enceladus and Europa, these oceans are in contact with the solid core, which may support ongoing metabolic processes such as methanogenesis and sulfate reduction \citep{zolotov2001composition,moran2008methyl,mckay2008possible,hsu2015ongoing,sekine2015high,waite2017cassini,weber2023review}, making them prime targets in the search for life beyond Earth. 

Direct measurements of icy moon oceans are impossible because they are hidden beneath thick ice shells. We therefore rely on indirect surface observations to infer the ocean circulation. One such observable is the ice thickness distribution \citep{vcadek2016enceladus,tajeddine2017true,vcadek2019long,hemingway2019enceladus,park2024global}. If the ice shell is in a quasi-equilibrium state, melting and freezing at the ice-ocean interface must balance the mass transport by the ice flow. The ice flow is driven by gradients in ice thickness \citep{vcadek2019long,shibley2024infer}, so these gradients can reveal the underlying ocean circulation through the heat flux at the ice-ocean interface, which controls the local freezing and melting rates \citep{zhang2025does,kang2026subsurface}.

\begin{figure}
    \centering
    \includegraphics[width=0.8\textwidth]{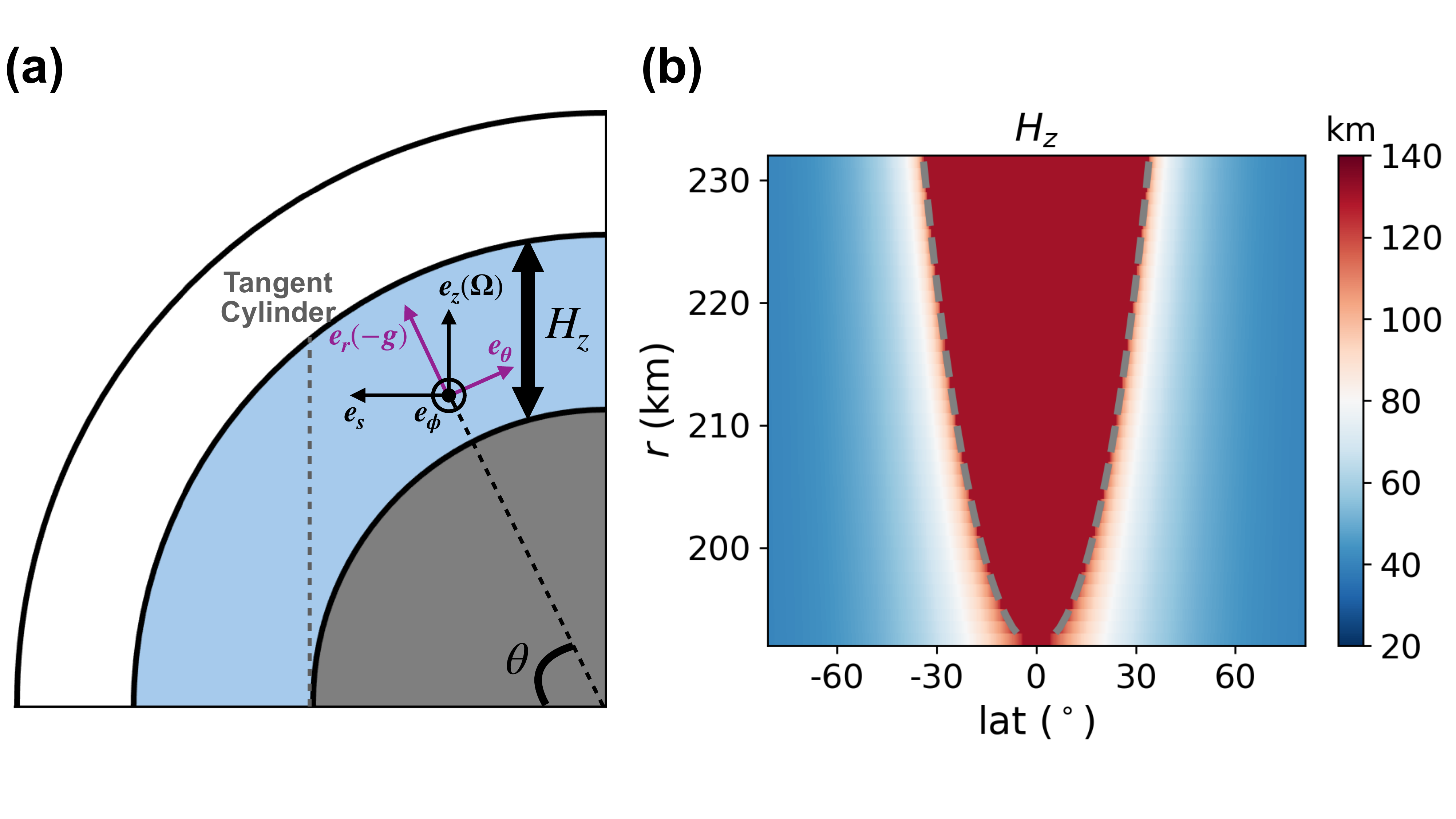}
    \caption{
    \textbf{Coordinate systems.} (a) Schematic of the coordinate systems. $\mathbf{\ey}$ and $\mathbf{e_r}$ are the meridional (latitudinal) direction and the vertical direction (opposite to the gravitational direction), respectively. $\mathbf{e_z}$ and $\mathbf{e_s}$ are the axial direction aligned with planetary rotation and the radial direction perpendicular to it, respectively. $\mathbf{e_\phi}$ represents both the zonal direction in spherical coordinates and the azimuthal direction in cylindrical coordinates. (b) Distribution of $H_z$ in the simulation with Enceladus-like parameters: a core radius of 192~km and an ocean surface radius of 232~km. Outside the tangent cylinder, $H_z$ is kept constant. In (a) and (b), gray dashed lines indicate the tangent cylinder.}
    \label{fig:coordinate}
\end{figure}

One of the dominant dynamical processes in icy moon oceans is thought to be convection driven by tidal heating in the silicate core \citep{tobie2005tidal,beuthe2013spatial,choblet2017powering}. Under strong rotational constraints, convection in icy moon oceans organizes into columnar structures aligned with the planetary rotation axis. The ocean can then be divided into two regions separated by the tangent cylinder, a virtual cylinder parallel to the rotation axis and tangent to the solid core (Fig.~\ref{fig:coordinate}). Inside the tangent cylinder, the bases of columns aligned with the rotation axis are in direct contact with the heated core, forming convective plumes known as slantwise convection. Outside the tangent cylinder, the columns are not directly connected to the core at their bases, and convection takes the form of equatorial rolls that circulate around the columns \citep{ashkenazy2021dynamic,bire2022exploring}. The relative importance of these modes in heat transport determines whether more heat is delivered to equatorial surfaces outside the tangent cylinder or to polar surfaces inside the tangent cylinder. This competition remains uncertain and can depend sensitively on parameter choices \citep{soderlund2019ocean,amit2020cooling,kvorka2022numerical,bire2022exploring,hartmann2024toward}. For example, \cite{bire2022exploring} showed that if the viscosity ($\nu$) is set too high, leading to a high Ekman number ($\mathrm{Ek} = \nu\Omega^{-1} H^{-2}$ where $\Omega$ is the planetary rotation rate and $H$ is the ocean depth), high-latitude convective plumes can be artificially suppressed. In more realistic low-viscosity regimes, they found enhancement of vertical heat transport toward the poles under uniform bottom heating, indicating that slantwise convection is more efficient at transporting heat than equatorial rolls.

\begin{table}
\centering
\caption{Icy moon ocean parameters. This table is adapted from Table~S1 in \protect\cite{zeng2026slantwise}.}
\label{tab:moons}
\begin{tabular}{c|cccc}
\hline
Parameter & Enceladus & Titan & Europa & Ganymede \\
\hline
$Q$ (mW~m$^{-2}$) & 16--83 & 14--20 & 23--123 & 15--107 \\
$g$ (m~s$^{-2}$) & 0.1 & 1.4 & 1.3 & 1.4 \\
$\alpha_T$ (10$^{-4}~$K$^{-1}$) & 0.1--1.3 & 0.4--4.2 & 1.9--2.5 & 1.9--4.4 \\
$2\Omega$ (s$^{-1}$) & $1.1 \times 10^{-4}$ & $9.2 \times 10^{-6}$ & $4.2 \times 10^{-5}$ & $2.0 \times 10^{-5}$ \\
$H$ (km) & 11--63 & 91--420 & 97--131 & 24--518 \\
\hline
$B$ ($10^{-14}$ m$^2$~s$^{-3}$) & 0.4--27.0 & 19.6--294 & 148--938 & 120--1650 \\
$l_{\mathrm{plume}}$ ($^\circ$) & 0.006--0.039 & 0.027--0.113 & 0.024--0.044 & 0.026--0.106 \\
$\mathrm{Ro}^*$ & $1\times 10^{-6}$--$4 \times 10^{-5}$ & $4\times 10^{-5}$--$5 \times 10^{-4}$ & $4\times 10^{-5}$--$1 \times 10^{-4}$ & $2\times 10^{-5}$--$4 \times 10^{-4}$ \\
% $\text{Ek}$ & $10^{-11}$--$10^{-10}$ & $10^{-12}$--$10^{-10}$ & $10^{-11}$ & $10^{-12}$--$10^{-11}$ \\
% $\text{Ra}$ & $10^{17}$--$10^{20}$ & $10^{19}$--$10^{24}$ & $10^{20}$--$10^{22}$ & $10^{20}$--$10^{24}$  \\
\hline
\end{tabular}
\end{table}

A major challenge for global simulations for icy moon oceans is that convection cannot be resolved explicitly. Rotating convection scaling \citep{fernando1989turbulent,jones1993convection} suggests that convective plumes in icy moon oceans are extremely small, about 0.01-0.1$^\circ$ in latitude (Table~\ref{tab:moons}). Two main approaches have been used to address this problem. The first approach performs simulations in a parameter regime where convection is resolvable and then extrapolates the results to real oceans using scaling laws \citep{soderlund2014ocean,soderlund2019ocean,amit2020cooling,kvorka2022numerical,bire2022exploring,lemasquerier2023europa}. However, these regimes differ substantially from those relevant to icy moon oceans, so the extrapolated results should be interpreted with caution. Moreover, most existing scaling laws for rotating convection focus on the vertical heat transport along the gravitational direction \citep{gastine2016scaling,wang_diffusionfree_2021,gastine2023latitudinal,hartmann2024toward,fan2024scaling,kannan2026scaling}, while quantitative scalings for meridional transport remain limited. \cite{zeng2026slantwise} derived scaling laws for both vertical and meridional heat transport of convection from local Cartesian simulations, but their applicability in global models remains to be examined.

The second approach parameterizes convective transports while resolving larger-scale dynamics such as baroclinic eddies and overturning circulations \citep{kang2022does,kang2023modulation,zeng2024effect}. Simulations with parameterized convection indicate that, for Enceladus, baroclinic eddies driven by surface buoyancy gradients \citep{vallis2017atmospheric} dominate heat transport over convection, leading to equatorward heat convergence that cannot maintain the observed poleward-thinning ice geometry. However, the convection scheme used in these studies was developed for Earth's ocean and assumes instability and adjustment along the gravitational direction \citep{marotzke1991influence}. In icy moon oceans, the natural Rossby number is small, $\Ron \equiv B^{1/2} (2\Omega)^{-3/2}H^{-1} \ll 1$, where $B$ is the buoyancy flux \citep{jones1993convection,maxworthy1994unsteady,bire2022exploring}. This indicates that convection is strongly constrained by rotation and tends to align with the rotation axis. This tilted structure implies that both the instability criterion and the transport direction differ from those assumed in the traditional convection scheme, raising questions about its applicability to icy moon oceans. In particular, traditional schemes do not represent the meridional heat transport associated with slantwise convection \citep{zeng2026slantwise}. A new slantwise convection scheme is therefore needed to accurately simulate ocean circulation on icy moons.

In this study, we develop a slantwise convection scheme based on the convection-resolving simulations of \cite{zeng2026slantwise} (hereafter \textit{ZJ26} simulations). This scheme captures the meridional transport by slantwise convection, which is neglected in the traditional convection scheme. The scheme is implemented in the Massachusetts Institute of Technology General Circulation Model \citep[MITgcm,][]{adcroft2018mitgcm}, a numerical model widely used to simulate icy moon oceans \citep[e.g.,][]{zeng2021ocean,kang2022does,kang2023modulation,zeng2024effect,ames2025ocean}. We conduct benchmark tests in the low-$\Ron$ regime relevant to icy moon oceans. Section~\ref{sec:scheme_development} describes the formulation of the slantwise convection scheme. Section~\ref{sec:benchmark} presents benchmark simulations in the low-$\Ron$ regime and compares the parameterized results with convection-resolving simulations in a global spherical shell. Section~\ref{sec:discussion} provides discussion and concluding remarks.

\section{Formulation of the scheme}\label{sec:scheme_development}

\subsection{Governing equations and coordinates}\label{subsec:equation}

We solve the Navier-Stokes equations under the Boussinesq approximation for icy moon oceans \citep{vallis2017atmospheric}:

\begin{equation}\label{eq:momemtum}
    \Dt{\bV} + 2 \bOmega \times \bV = -\nabla \Phi + b \eR + \nabla \cdot ( \boldsymbol{\nu} \nabla \bV),
\end{equation}

\begin{equation}\label{eq:temperature}
    \Dt{T} = \nabla \cdot (\boldsymbol{\kappa} \nabla T),
\end{equation}

\begin{equation}\label{eq:salinity}
    \Dt{S} = \nabla \cdot (\boldsymbol{\kappa} \nabla S),
\end{equation}

\begin{equation}\label{eq:continuity}
    \nabla \cdot \bV = 0,
\end{equation}

\noindent where $\bV$ is the velocity vector, $\nabla \Phi$ is the pressure gradient term, $b=b(S,T)$ is the buoyancy, $S$ is the salinity, and $T$ is the temperature. $\boldsymbol{\nu}$ and $\boldsymbol{\kappa}$ are the three-dimensional (3-D) eddy viscosity and diffusivity tensors representing unresolved subgrid-scale mixing processes. We use $(\boldsymbol{e_\phi},\boldsymbol{\ey},\boldsymbol{e_r})$ to denote the zonal, meridional, and vertical directions in spherical coordinates, where $\phi$ is longitude, $\theta$ is latitude, and $r$ is the spherical radius, and $(\boldsymbol{e_\phi},\boldsymbol{e_z},\boldsymbol{e_s})$ to denote the azimuthal, axial, and radial directions in cylindrical coordinates, respectively (Fig.~\ref{fig:coordinate}a; note that the zonal direction in spherical coordinates and the azimuthal direction in cylindrical coordinates are the same).

\subsection{From traditional upright convection to slantwise convection}\label{subsec:scaling}

To incorporate slantwise convection, two main aspects of the traditional convection scheme widely used for Earth's ocean \citep{marotzke1991influence} must be modified. The first is the instability criterion. The traditional scheme is triggered when stratification is unstable along the gravitational direction, i.e., $\partial b/\partial r < 0$. This criterion is appropriate for upright convection on Earth. However, in icy moon oceans, the Rossby number is small, so angular momentum is dominated by planetary rotation, and surfaces of constant angular momentum are approximately parallel to the rotation axis. In this regime, the flow is always inertially stable, and the only possible buoyancy-driven instability is governed by stratification along the rotation axis. Instability therefore occurs if $(\partial b/\partial z) \si<0$, where the factor $\si$ accounts for the projection of gravity relative to the rotation axis \citep{zeng2025symmetric}.

The second aspect is the representation of convective buoyancy transports. The traditional scheme parameterizes convection using a large vertical diffusivity, $\kappa_{rr,c}$, typically prescribed as a constant, where the subscript $c$ here denotes the convective contribution to the eddy diffusivity. This formulation represents only upright transport and adjustment of the buoyancy profile along the gravitational direction, while neglecting any meridional fluxes associated with the tilted structure of slantwise convection. Moreover, it does not depend on local buoyancy gradients, which control the strength of convection.

\begin{figure}
    \centering
    \includegraphics[width=0.9\textwidth]{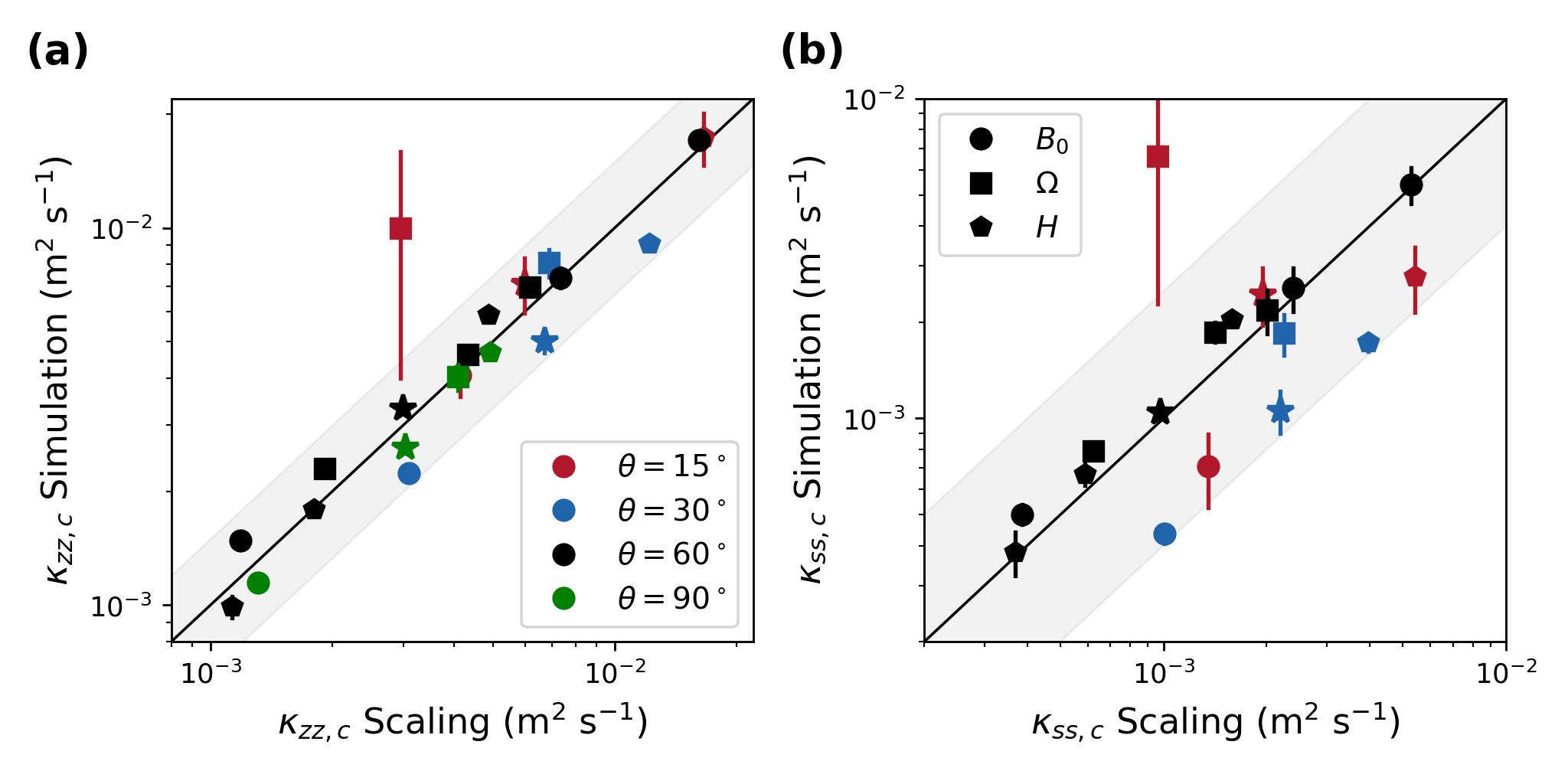}
    \caption{\textbf{Scalings of diffusivities.} Panels (a) and (b) compare the diffusivity scalings in the axial and radial directions (Equations~\ref{eq:kappazz} and \ref{eq:kappass}) with convection-resolving simulations from \protect\cite{zeng2026slantwise}, respectively. Marker color denotes latitude (legends in a), and shape denotes parameter variations (legends in b). Black lines denote the one-to-one line. Gray shading indicates the range between $0.67\times$ and $1.5\times$ the scaling in panel (a), and between $0.4\times$ and $2.5\times$ in panel (b). Most simulations lie within this range, except for one case exhibiting large temporal variability (see \protect\cite{zeng2026slantwise} for details).}
    \label{fig:kappa_scaling}
\end{figure}

To address these limitations, we represent convective transport using an anisotropic diffusivity tensor $\boldsymbol{\kappa_c}$ derived from \textit{ZJ26} simulations. The buoyancy flux along the rotation axis follows a CIA (Coriolis--Inertial--Archimedean) scaling \citep{aurnou2020connections}:

\begin{equation}\label{eq:kappazz}
    B_z = -\kappa_{zz,c} \frac{\partial b}{\partial z}, \ \
    \kappa_{zz,c} =
    \begin{cases}
        K_z \left|\dfrac{\partial b}{\partial z}\right|^{3/2}
        (2\Omega)^{-2} H_z^{2},
        & \mathrm{if} \ \dfrac{\partial b}{\partial z}\sin\theta < 0 \ \ \mathrm{(Unstable)},\\
        0,
        & \mathrm{if} \ \dfrac{\partial b}{\partial z}\sin\theta \ge 0 \ \ \mathrm{(Stable)}.
    \end{cases}
\end{equation}

\noindent where $H_z$ is the characteristic depth scale of convection plumes and $K_z=0.660$ is an empirical constant. This scaling reproduces the \textit{ZJ26} simulation results well (Fig.~\ref{fig:kappa_scaling}a). 

Inside the tangent cylinder of a spherical shell, the depth scale of slantwise convection plumes is constrained by the shell geometry and therefore varies with latitude, assuming that convection spans the full ocean depth inside the tangent cylinder. Outside the tangent cylinder, however, the dynamics are dominated by equatorial rolls rather than slantwise convection, making the appropriate choice of $H_z$ less clear. In the present formulation, we keep $H_z$ constant outside the tangent cylinder (Fig.~\ref{fig:coordinate}). We also tested alternative formulations in which $H_z$ is constrained by the spherical boundaries outside the tangent cylinder, but found no qualitative changes in the results (Appendix~\ref{app:Hz}).

For the radial direction, we assume that the diffusivity is proportional to that in the axial direction but reduced by a constant factor:

\begin{equation}\label{eq:kappass}
    B_s = -\kappa_{ss,c} \frac{\partial b}{\partial s}, \ \kappa_{ss,c} = \gamma \kappa_{zz,c},
\end{equation}

\noindent where $\gamma = 0.326$ is an empirical constant obtained from the simulations of \textit{ZJ26}. With this factor, the scaling agrees with the simulations to within a factor of 2.5 (Fig.~\ref{fig:kappa_scaling}b). The diffusivity $\kappa_{zz,c}$ represents transport associated with slantwise convective motions along the rotation axis, while $\kappa_{ss,c}$ represents cross-column transport. Because these processes are dynamically distinct, we represent the diffusivity tensor as anisotropic but diagonal in cylindrical coordinates (i.e., $\kappa_{zs,c}=\kappa_{sz,c}=0$). %The result $\gamma<1$ indicates that heat transport is more efficient along the rotation axis than across columns, consistent with results from \cite{bire2022exploring}, which suggest that slantwise convection is more efficient than equatorial rolls under similar conditions.

\subsection{Tensor transformation}\label{subsec:tensor}

General circulation models for global icy moon ocean simulations are formulated in spherical coordinates to accommodate realistic boundary conditions. We therefore transform the diffusivity tensor from cylindrical coordinates to spherical coordinates. We focus on the meridional heat transport and therefore only require the transformation between the $(\boldsymbol{e_z},\boldsymbol{e_s})$ and $(\boldsymbol{e_\theta},\boldsymbol{e_r})$ directions:

\begin{equation}
    \boldsymbol{\kappa_c}=\begin{pmatrix} \kappa_{\theta \theta,c} & \kappa_{\theta r,c} \\ \kappa_{r \theta,c} & \kappa_{rr,c} \\ \end{pmatrix} = 
    \boldsymbol{R}^T
    \begin{pmatrix} \kappa_{zz,c} & 0 \\ 0 & \kappa_{ss,c} \end{pmatrix}
    \boldsymbol{R} = \kappa_{zz,c}
    \begin{pmatrix} \coe{2}+\gamma \sie{2} & (1-\gamma) \si \co \\ (1-\gamma) \si \co & \sie{2}+\gamma \coe{2} \end{pmatrix},
\end{equation}

\noindent where

\begin{equation*}
    \boldsymbol{R} = \begin{pmatrix} \co & \si \\ -\si & \co \end{pmatrix}
\end{equation*}

\noindent is the rotation matrix. The slantwise convection diffusivity is added to the background eddy diffusivity to form the total eddy diffusivity, $\boldsymbol{\kappa}=\boldsymbol{\kappa_b}+\boldsymbol{\kappa_c}$. The slantwise convection scheme does not modify the eddy viscosity, so that $\boldsymbol{\nu}=\boldsymbol{\nu_b}$. Here, the subscript $b$ denotes the background contribution to the eddy diffusivity and viscosity. The scheme is implemented in MITgcm, which uses an Arakawa-C grid \citep{arakawa1977computational}.

\section{Benchmark tests}\label{sec:benchmark}

\subsection{Model configuration}\label{subsec:model}

The scaling laws in Section~\ref{subsec:scaling} are derived from local Cartesian simulations. To assess their applicability in a more realistic geometry, we conduct convection-resolving simulations (CR simulations) in a global spherical shell using MITgcm. In addition, we perform simulations with the slantwise convection parameterization (Equations~\ref{eq:kappazz}~\&~\ref{eq:kappass}) under the same setup for comparison (SL simulations). These simulations serve as benchmark tests for the parameterization developed in this study. We also perform simulations with the traditional convection scheme \citep[``convective adjustment scheme'' in MITgcm,][]{adcroft2018mitgcm} for comparison (TR simulations).

We impose a uniform bottom heat flux $Q_{bot}$ and restore the surface temperature toward the freezing point (0$^\circ$C) at a time scale $\tau_T$ to represent icy moon conditions. The uniform surface temperature suppresses lateral buoyancy gradients and therefore minimizes the influence of baroclinic eddies, allowing us to isolate the effect of slantwise convection. We fix the ocean salinity and adopt a linear equation of state with thermal expansion coefficient $\alpha_T = 2\times10^{-4}\ \mathrm{K^{-1}}$, such that $b = -g\alpha_T(T-T_{ref})$, where we choose $T_{ref} = 0^\circ\mathrm{C}$. Other parameters are chosen to be representative of Enceladus: rotation rate $\Omega = 5\times10^{-5}\ \mathrm{s^{-1}}$, gravity $g = 0.1\ \mathrm{m\,s^{-2}}$, bottom radius $r_{bot} = 192\ \mathrm{km}$, and surface radius $r_{surf} = 232\ \mathrm{km}$. The resulting $H_z$ profile is shown in Fig.~\ref{fig:coordinate}b.

\begin{table}
\caption{\textbf{Simulation setup.} CR indicates convection-resolving simulations, SL indicates simulations with the slantwise convection parameterization, and TR indicates the simulation with the traditional convection scheme. The restoring timescale $\tau_T$ is chosen to be approximately 1/50 of the vertical overturning timescale. In the TR simulations, we set $\kappa_{rr,c}=3869$, 972, 244, and 61.3~m$^2$~s$^{-1}$ for simulations from the highest to lowest heat fluxes, respectively. These values are chosen to be comparable to the diffusivities estimated from the corresponding SL simulations.}
\centering
\begin{tabular}{ccccccc}
\toprule
Scheme & $Q_{bot}$ & $Ro^*$ & $\mathrm{Ek}= \nu \Omega^{-1} H^{-2}$ & Resolution & $\tau_T$ & $\tau_U$ \\
\midrule
\multirow{4}{*}{CR}
& $10^6$ W~m$^{-2}$ & $5.6 \times 10^{-2}$ & $5.8 \times 10^{-4}$ & $0.45^\circ$ & $10^4$~s & $10^5$~s\\
& $10^5$ W~m$^{-2}$ & $1.8 \times 10^{-2}$ & $2.7 \times 10^{-4}$ & $0.45^\circ$ & $3 \times 10^4$~s & $10^5$~s\\
& $10^4$ W~m$^{-2}$ & $5.6 \times 10^{-3}$ & $1.3 \times 10^{-4}$ & $0.225^\circ$ & $9 \times 10^4$~s & $10^5$~s\\
& $10^3$ W~m$^{-2}$ & $1.8 \times 10^{-3}$ & $5.8 \times 10^{-5}$ & $0.225^\circ$ & $2.7 \times 10^5$~s & $10^6$~s\\
\midrule
\multirow{4}{*}{SL}
& $10^6$ W~m$^{-2}$ & $5.6 \times 10^{-2}$ & $3.8 \times 10^{-3}$ & $0.9^\circ$ & $10^4$~s & $10^5$~s\\
& $10^5$ W~m$^{-2}$ & $1.8 \times 10^{-2}$ & $1.7 \times 10^{-3}$ & $0.9^\circ$ & $3 \times 10^4$~s & $10^5$~s\\
& $10^4$ W~m$^{-2}$ & $5.6 \times 10^{-3}$ & $8.1 \times 10^{-4}$ & $0.9^\circ$ & $9 \times 10^4$~s & $10^5$~s\\
& $10^3$ W~m$^{-2}$ & $1.8 \times 10^{-3}$ & $3.8 \times 10^{-4}$ & $0.9^\circ$ & $2.7 \times 10^5$~s & $10^6$~s\\
\midrule
\midrule
\multirow{4}{*}{TR}
& $10^6$ W~m$^{-2}$ & $5.6 \times 10^{-2}$ & $3.8 \times 10^{-3}$ & $0.9^\circ$ & $10^4$~s & $10^5$~s\\
& $10^5$ W~m$^{-2}$ & $1.8 \times 10^{-2}$ & $1.7 \times 10^{-3}$ & $0.9^\circ$ & $3 \times 10^4$~s & $10^5$~s\\
& $10^4$ W~m$^{-2}$ & $5.6 \times 10^{-3}$ & $8.1 \times 10^{-4}$ & $0.9^\circ$ & $9 \times 10^4$~s & $10^5$~s\\
& $10^3$ W~m$^{-2}$ & $1.8 \times 10^{-3}$ & $3.8 \times 10^{-4}$ & $0.9^\circ$ & $2.7 \times 10^5$~s & $10^6$~s\\
\midrule
\multirow{4}{*}{SL}
& $10^4$ W~m$^{-2}$ & $5.6 \times 10^{-3}$ & $3.2 \times 10^{-4}$ & $0.45^\circ$ & $9 \times 10^4$~s & $10^5$~s\\
& $10^4$ W~m$^{-2}$ & $5.6 \times 10^{-3}$ & $1.4 \times 10^{-3}$ & $1.35^\circ$ & $9 \times 10^4$~s & $10^5$~s\\
& $10^4$ W~m$^{-2}$ & $5.6 \times 10^{-3}$ & $2.4 \times 10^{-3}$ & $0.9^\circ$ & $9 \times 10^4$~s & $10^5$~s\\
& $10^4$ W~m$^{-2}$ & $5.6 \times 10^{-3}$ & $2.7 \times 10^{-4}$ & $0.9^\circ$ & $9 \times 10^4$~s & $10^5$~s\\
\bottomrule
\end{tabular}
\label{tab:simulation_setup}
\end{table}

Resolving slantwise convection with realistic heat fluxes is computationally infeasible because the resulting convective plumes are too small (Table~\ref{tab:moons}). We therefore apply enhanced heat fluxes to make the plumes resolvable, while keeping a small natural Rossby number: $\Ron = 5.6 \times 10^{-2}, \ 1.8 \times 10^{-2}, \ 5.6 \times 10^{-3}, \ $ and $1.8 \times 10^{-3}$. For reference, we also diagnose the horizontally averaged vertical temperature profile to estimate the Rayleigh number ($\mathrm{Ra}=g\alpha_T \Delta T H^3/(\nu \kappa)$) for each CR simulation (see Table~\ref{tab:Ra_Ek} in Appendix~\ref{app:Ra_Ek}).

In all simulations, we apply isotropic background eddy diffusivity and viscosity and fix the Prandtl number $\mathrm{Pr} = \nu_b/\kappa_b = 10$ to maintain numerical stability. The vertical resolution is 1000~m. For the CR simulations, the horizontal resolution is $0.45^\circ$ for $\Ron = 5.6 \times 10^{-2}$ and $1.8 \times 10^{-2}$, and $0.225^\circ$ for $\Ron = 5.6 \times 10^{-3}$ and $1.8 \times 10^{-3}$, in order to resolve the convective plumes (see Appendix~\ref{app:Ra_Ek} for more details). The SL (slantwise convection parameterization) and TR (traditional convection parameterization) simulations use a coarser horizontal resolution of $0.9^\circ$ with increased viscosity, since convection is parameterized rather than explicitly resolved. 

To assess sensitivity to numerical parameters, we additionally perform SL simulations with different horizontal resolutions ($0.45^\circ$ and $1.35^\circ$)\footnote{For simulations with varying horizontal resolution, the viscosity is also adjusted according to Kolmogorov scaling, $\nu \propto L_{grid}^{4/3}$ \citep[cf. Appendix~B in][]{zeng2021ocean}.} and viscosities ($3\nu$ and $\nu/3$), while keeping $\mathrm{Pr} = 10$. To reduce computational cost, the zonal domain is restricted to $13.5^\circ$ with periodic boundary conditions, corresponding to an approximate 27-fold zonal symmetry. The meridional extent spans $81^\circ$S to $81^\circ$N to avoid numerical difficulties near the poles, where zonal grid spacing becomes small. We apply no-normal-flow and free-slip boundaries at the top and bottom. To prevent long-term drift of the zonal flow, we apply a linear bottom drag, which restores the velocity in the lowest model level toward zero with a timescale $\tau_U=10^{5}$--$10^6$~s. All simulations are integrated to a statistically steady state, and time-averaged results are presented to reduce temporal variability, unless otherwise noted. Detailed simulation parameters are listed in Table~\ref{tab:simulation_setup}.

\begin{figure}
    \centering
    \includegraphics[width=\textwidth]{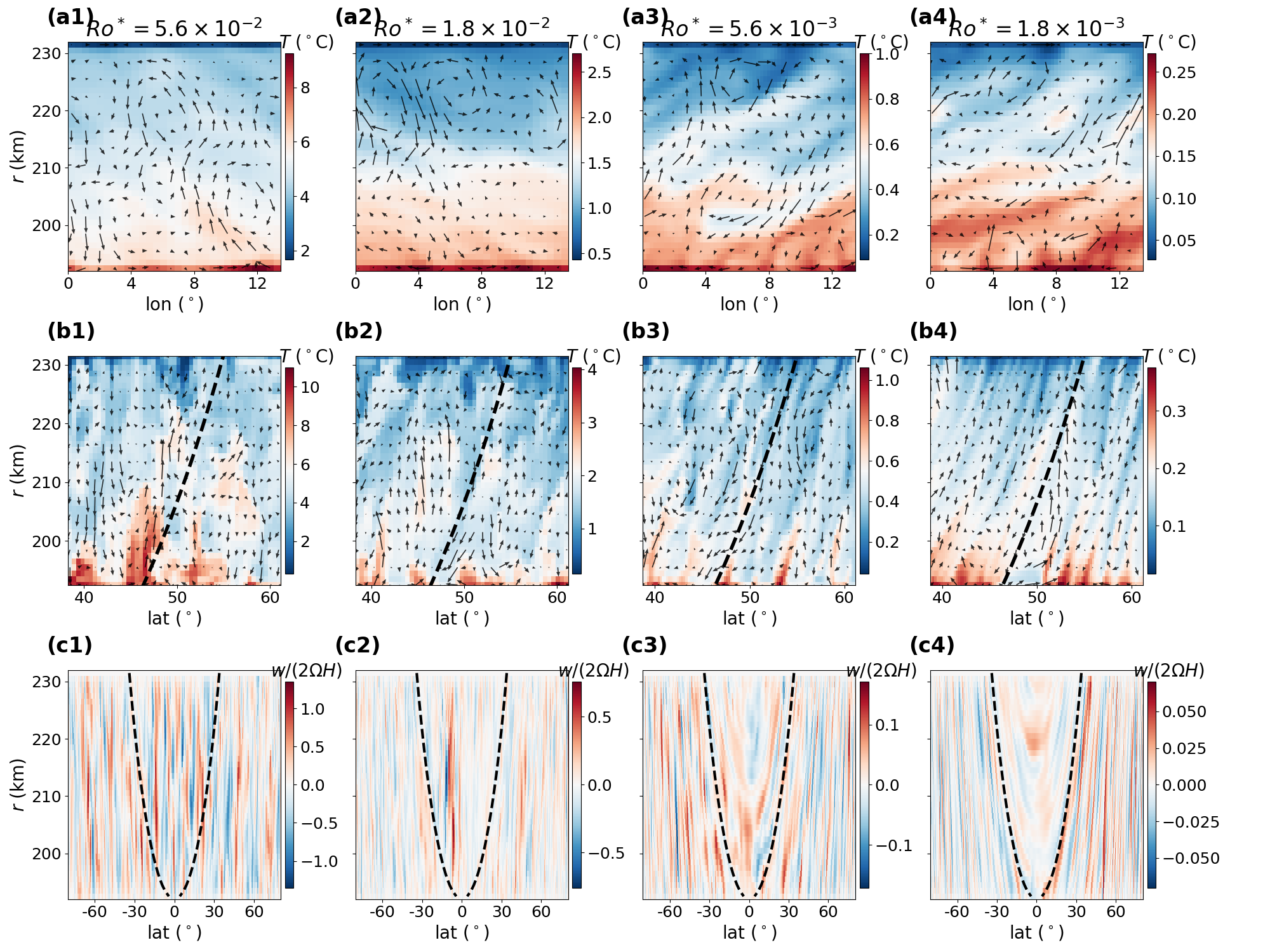}
    \caption{\textbf{Snapshots of convective plumes in convection-resolving (CR) simulations.} (a): Snapshots of equatorial rolls, longitude-depth sections averaged over a $10^\circ$ band around the equator. (b): Snapshots of slantwise convection at mid-latitudes, latitude-depth sections for one longitude slice. Color shadings indicate temperature and arrows indicate velocity anomalies with the zonal mean subtracted. In (b), dashed lines indicate the direction parallel to rotation ($s=r\cos{\theta}=\mathrm{const}$). (c): Snapshots of vertical velocity $w$ at one longitude slice. The velocity is non-dimensionalized by rotation rate $\Omega$ and ocean depth $H$, representing the Rossby number of the convective flow. Black dashed lines indicate the tangent cylinder. From left to right, the columns correspond to convection-resolving simulations with $\Ron = 5.6 \times 10^{-2}, \ 1.8 \times 10^{-2}, \ 5.6 \times 10^{-3},$ and $1.8 \times 10^{-3}$, respectively.}
    \label{fig:high-res_snaps}
\end{figure}

\begin{figure}
    \centering
    \includegraphics[width=\textwidth]{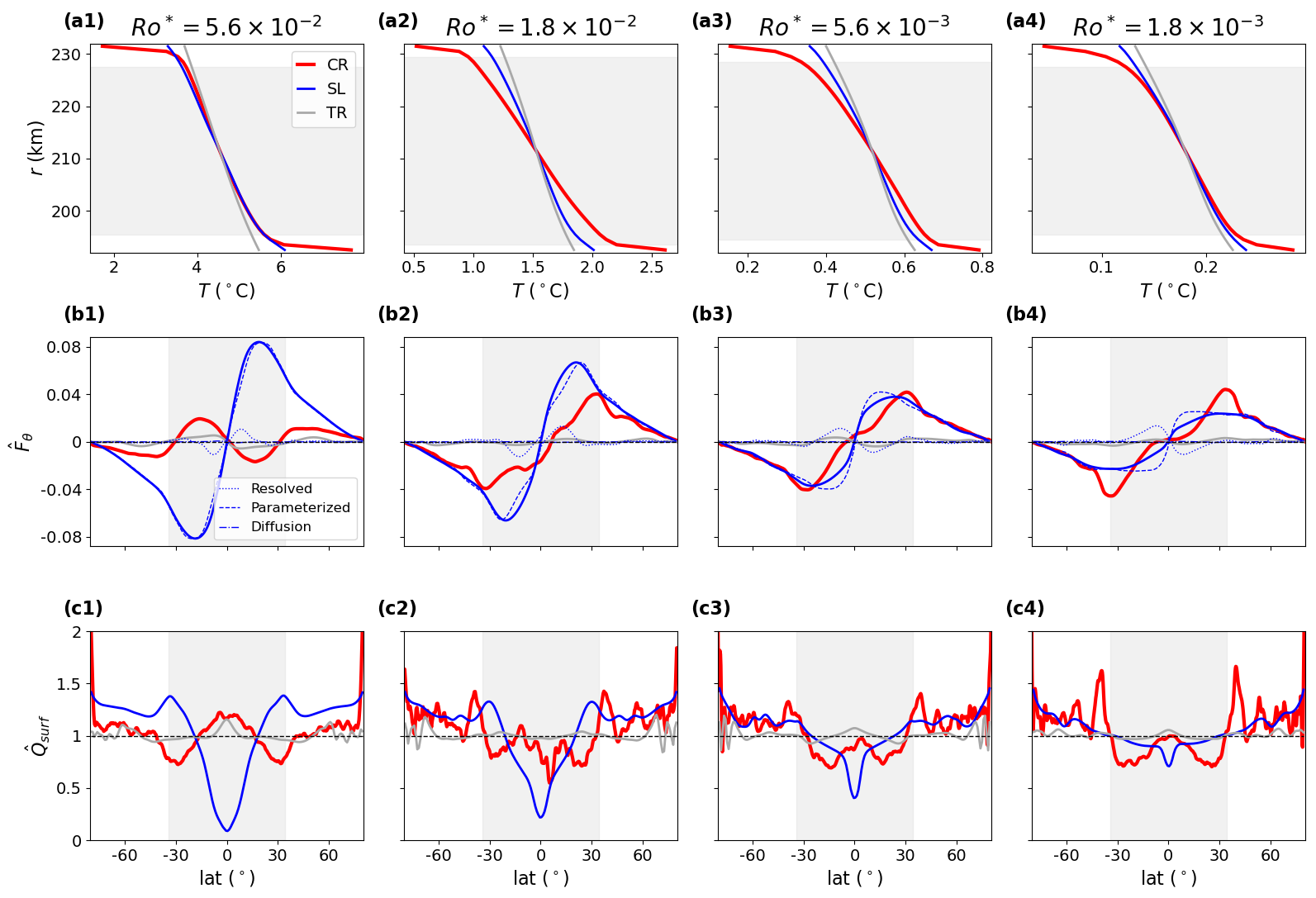}
    \caption{\textbf{Benchmark test results.} (a): Horizontally averaged temperature profiles. (b): Vertically integrated meridional heat flux $F_\theta$, normalized by the total bottom heat flux as $\hat{F}_\theta = F_\theta/(Q_{bot}A_{bot})$, where $A_{bot}$ is the total bottom area. (c) Surface heat flux per unit area $Q_{surf}$, normalized by the bottom heat flux rescaled to the surface as $\hat{Q}_{surf} = (Q_{surf}A_{surf})/(Q_{bot}A_{bot})$, where $A_{surf}$ is the total surface area. Red lines indicate convection-resolving (CR) simulations, blue lines indicate simulations with the slantwise convection parameterization developed in this study (SL), and gray lines indicate simulations using the traditional convection scheme (TR). In (a), gray shading indicates the interior region, defined as the region where the vertical temperature profile is approximately linear (see \protect\cite{zeng2026slantwise} for details of the algorithm separating interior and boundary layers). Blue and gray lines are shifted by a constant offset for better comparison of the interior temperature gradient. In (b), blue dotted, dashed, and dash-dotted lines indicate contributions from resolved, parameterized, and background diffusive fluxes, respectively. In (b) and (c), gray shading indicates regions outside the tangent cylinder.}
    \label{fig:benchmark_heat}
\end{figure}

\subsection{Simulation results}\label{subsec:results}

In convection-resolving (CR) simulations, we observe distinct dynamical features in equatorial rolls (Fig.~\ref{fig:high-res_snaps}a) and mid-latitude slantwise convection (Fig.~\ref{fig:high-res_snaps}b) across different $\Ron$. In the smaller-$\Ron$ simulations ($\Ron=5.6 \times 10^{-3}$ and $1.8 \times 10^{-3}$), mid-latitude convection aligns with the rotation axis, consistent with the rapidly rotating regime \citep{aurnou2020connections}. In the equatorial regions we see indications for eastward tilted rolls, resembling equatorial Busse modes associated with superrotation, i.e., an eastward jet at the equator \citep{busse1982differential}. In contrast, in the larger-$\Ron$ simulations ($\Ron=5.6 \times 10^{-2}$ and $1.8 \times 10^{-2}$), the equatorial rolls tend to tilt westward and the mid-latitude convection no longer exhibits clear rotational alignment. These differences are also evident in the vertical velocity fields (Fig.~\ref{fig:high-res_snaps}c), indicating that the larger-$\Ron$ cases lie outside the rapidly rotating regime relevant to icy moon oceans. This transition is broadly consistent with \cite{bire2022exploring}, who found that rotationally constrained convection occurs approximately for $\Ron<10^{-2}$. Realistic icy moon oceans typically have $\Ron \sim 10^{-6}$--$10^{-4}$ (Table~\ref{tab:moons}), substantially smaller than the values explored here.

For the smaller-$\Ron$ cases, simulations with the slantwise convection parameterization (SL) reproduce the total meridional heat transport in the CR simulations well (Fig.~\ref{fig:benchmark_heat}b3~\&~b4). In particular, inside the tangent cylinder (i.e. at mid- to high-latitudes), the parameterized flux matches the total meridional heat transport in the CR simulations (compare dashed blue and solid red lines), while the resolved flux (blue dotted lines) and background diffusive flux (blue dash-dotted lines) are negligible. Outside the tangent cylinder, the agreement between the parameterized and CR fluxes becomes weaker, consistent with the fact that the low latitudes are not dominated by slantwise convection. In the two simulations with larger $\Ron$, the parameterized meridional heat transport deviates substantially from the CR results, as these cases are not within the rotation-constrained slantwise convection regime assumed in the parameterization.

The meridional heat transport significantly modifies the surface heat flux distribution. In all simulations, it modulates the surface heat flux by up to a factor of 2 relative to the uniform bottom heating (Fig.~\ref{fig:benchmark_heat}c). In contrast, simulations using the traditional convection scheme (gray lines in Fig.~\ref{fig:benchmark_heat}) exhibit negligible meridional heat transport, resulting in a nearly uniform surface heat flux. This leads to an underestimation of polar heat accumulation for a uniform surface temperature, which in turn could affect predictions of ice shell thickness distribution.

\begin{figure}
    \centering
    \includegraphics[width=\textwidth]{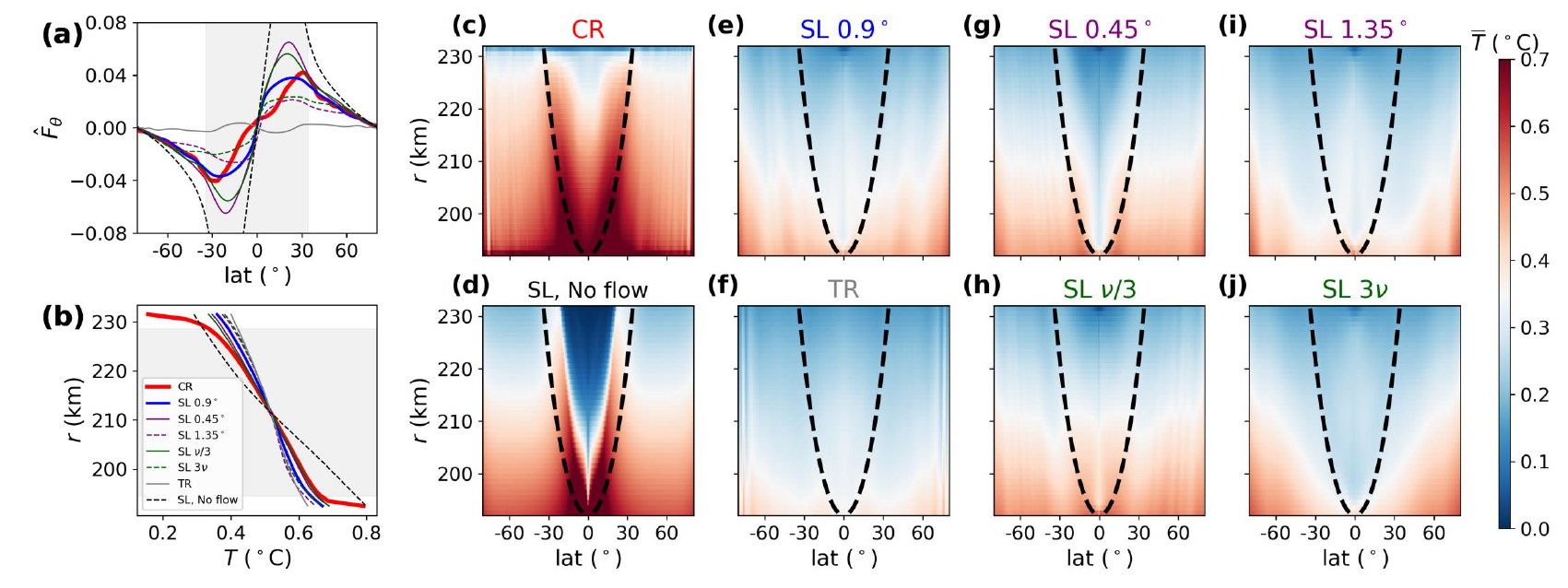}
    \caption{\textbf{Heat transport and temperature fields in the sensitivity tests with $\Ron=5.6 \times 10^{-3}$.} (a) and (b) are the same as Fig.~\ref{fig:benchmark_heat}b3 and a3, respectively, but include sensitivity tests. (c)-(j) show the zonal-mean temperature $\overline{T}$. Purple indicates sensitivity tests with varying horizontal resolution, and green indicates simulations with varying background viscosity. In (a), the gray shading indicates regions outside the tangent cylinder. In (b), the gray shading indicates the interior region in the CR simulation. In (c)-(i), black dashed lines indicate the tangent cylinder.}
    \label{fig:sensitivity:heat}
\end{figure}

Although the slantwise convection scheme reproduces the meridional heat transport well, the simulations with parameterized convection struggle to fully reproduce the temperature structure (Fig.~\ref{fig:benchmark_heat}a and Fig.~\ref{fig:sensitivity:heat}). In the SL simulations, there are no thermal boundary layers seen in the CR simulations, where large vertical temperature gradients occur. This is because the parameterization is derived for the ocean interior and does not represent boundary layers \citep{zeng2026slantwise}. The parameterized diffusivity is much larger than the background diffusivity, which smooths out sharp vertical gradients. Nevertheless, the interior temperature gradients in the SL simulations agree reasonably well with those in the CR simulations for the smaller-$\Ron$ cases (Fig.~\ref{fig:benchmark_heat}a3-a4), indicating that the scheme captures the bulk interior structure in the rotation-constrained regime.

Besides the missing boundary layers, differences also appear in the latitudinal distribution of temperature fields. In the CR simulation, there is a temperature maximum near the tangent cylinder (Fig.~\ref{fig:sensitivity:heat}c), associated with upwelling flow along the tangent cylinder (Fig.~\ref{fig:sensitivity:flow}b1). This feature is not reproduced in the SL simulation, and the overall temperature contrast in the SL simulation is weaker compared to the CR simulation (Fig.~\ref{fig:sensitivity:heat}e). This misfit is likely again attributable to the treatment of the boundary layers, since any mismatch in the meridional temperature structure at the outer edge of the boundary layer can project into the interior temperature field. However, it should also be noted that the boundary layers in our convection-resolving simulations are not expected to faithfully represent those in realistic icy moon oceans. Adequate parameterizations of the boundary layer dynamics are ultimately required, but developing such parameterizations is beyond the scope of this study.

\subsection{Sensitivity tests}\label{subsec:sensitivity}

To better understand the discrepancy in the temperature fields, we perform a series of sensitivity tests for the $\Ron=5.6 \times 10^{-3}$ case. This case lies within the rapidly rotating regime while allowing the convective plumes to be better resolved than in the simulation with the smallest $\Ron$. In the first experiment, the resolved ocean dynamics are turned off and all heat redistribution is represented solely by the slantwise convection scheme, together with a very weak background diffusion. In this case, the simulation reproduces the temperature maximum near the tangent cylinder, although the maximum shifts slightly equatorward (Fig.~\ref{fig:sensitivity:heat}d). This behavior is expected because heat transport is most efficient near the tangent cylinder. If all heat transport were aligned with the rotation axis, the temperature maximum would occur at the tangent cylinder, where the area projection from the bottom boundary onto columns parallel to the rotation axis is largest (cf. Appendix A in \citealt{zeng2021ocean}). In the parameterization, however, heat transport is not purely axial. The additional cross-column diffusivity shifts the effective transport direction slightly away from the rotation axis, causing the temperature maximum to move slightly equatorward. This simulation also overestimates the interior vertical temperature gradient and the meridional heat transport (Fig.~\ref{fig:sensitivity:heat}a~\&~b, black dashed lines). The likely explanation is that the scaling law underestimates the effective cross-column diffusivity outside the tangent cylinder, where the dynamics are dominated by equatorial rolls. A similar underestimation in the radial heat transport is also evident in the low-latitude simulations of \cite{zeng2026slantwise}. Near the equator, this cross-column diffusivity is approximately equivalent to the vertical diffusivity. Consequently, a larger temperature gradient is required to maintain the prescribed vertical heat flux. This enhanced temperature gradient then increases the heat transport along the rotation axis, which is approximately meridional near the equator. As a result, this simulation overestimates the meridional heat transport outside the tangent cylinder by approximately a factor of four.

The temperature field in the simulation with resolved dynamics turned off differs substantially from the simulation with active ocean dynamics, suggesting that the resolved dynamics play an important role in shaping the temperature structure and the associated heat transport even in the relatively coarse simulations with parameterized convection. To further examine this effect, we carry out sensitivity tests by varying the horizontal resolution and background viscosity. In the simulation with finer horizontal resolution (SL $0.45^\circ$) and the simulation with reduced viscosity (SL $\nu/3$), the resolved dynamics become stronger, and the heat transport outside the tangent cylinder increases by about 50\%. In contrast, in the simulation with coarser horizontal resolution (SL $1.35^\circ$) and the simulation with increased viscosity (SL $3\nu$), the resolved dynamics become weaker, and the meridional heat transport outside the tangent cylinder is reduced by about 50\%. Despite these differences, all SL simulations reproduce the heat transport inside the tangent cylinder reasonably well and capture the bulk interior vertical temperature gradient, while none of them reproduces the exact latitudinal temperature structure in the CR simulation (Fig.~\ref{fig:sensitivity:heat}). For comparison, the simulation with the traditional convection scheme (TR) exhibits even weaker temperature gradients and produces a very weak and largely wrong-signed meridional heat flux. This difference arises because the traditional parameterization applies strong diffusion only along the gravitational direction, without enhancing meridional diffusion.

\begin{figure}
    \centering
    \includegraphics[width=\textwidth]{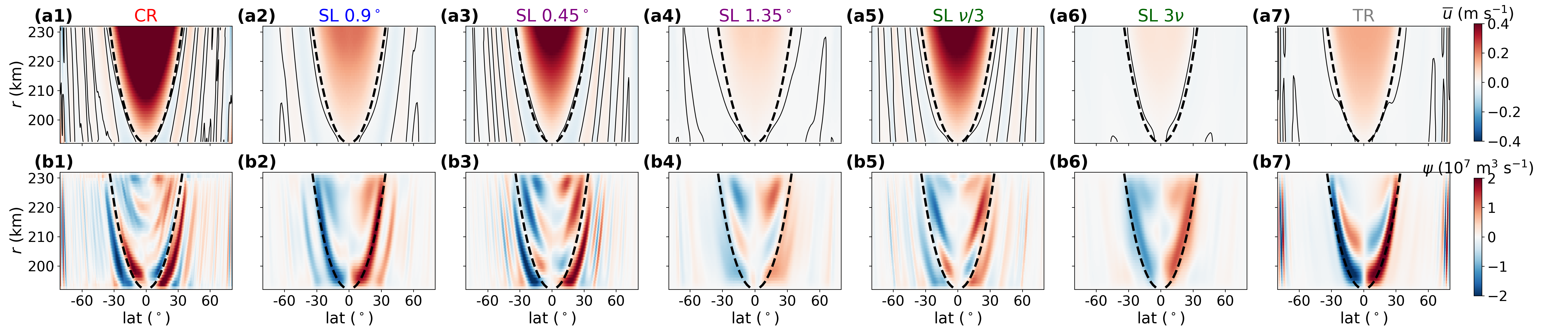}
    \caption{\textbf{Flow fields in the sensitivity tests with $\Ron=5.6 \times 10^{-3}$.} (a): zonal-mean zonal velocity $\overline{u}$. (b): stream function $\psi$, calculated as $\psi=-2 \pi \int_{r_{\mathrm{core}}}^{r_{\mathrm{surf}}} \overline{v} r \cos{\phi} \, d r$, where $\overline{v}$ is the zonal-mean meridional velocity, and $r_{\mathrm{core}}$ and $r_{\mathrm{surf}}$ are the radii of the core and ocean surface, respectively. Positive values of $\psi$ indicate clockwise circulation. Black dashed lines indicate the tangent cylinder. In (a), black solid lines indicate the $\overline{u}=0$ contours.}
    \label{fig:sensitivity:flow}
\end{figure}

Although the slantwise convection scheme does not include momentum transport, it is nevertheless interesting to examine its impact on the flow fields. In the CR simulation, a strong superrotating jet occupies the region outside the tangent cylinder, together with multiple alternating jets inside the tangent cylinder (Fig.~\ref{fig:sensitivity:flow}a1). In all SL simulations and the TR simulation, there is a superrotating jet outside the tangent cylinder but it is much weaker. These simulations also fail to reproduce the multiple alternating jets inside the tangent cylinder. Only the SL $0.45^\circ$ and SL $\nu/3$ simulations, which resolve more dynamical processes and hence can better represent the momentum fluxes, better reproduce the equatorial jet magnitude and the alternating jet structures (Fig.~\ref{fig:sensitivity:flow}a3~\&~a5).

The meridional overturning circulation also differs substantially between the simulations. In the CR simulation, multiple small-scale slantwise overturning cells appear inside the tangent cylinder, together with larger overturning cells outside the tangent cylinder. Near the tangent cylinder, a pair of overturning cells forms, with upwelling at the tangent cylinder and downwelling on its flanks (Fig.~\ref{fig:sensitivity:flow}b1). This structure is not reproduced in the SL and TR simulations. Many of them instead exhibit a single overturning cell spanning across the tangent cylinder (Fig.~\ref{fig:sensitivity:flow}b2-b7), which leads to equatorward convergence of heat by resolved advective transport (dotted blue line in Fig.~\ref{fig:benchmark_heat}b3).

\section{Discussion}\label{sec:discussion}

We developed a slantwise convection scheme to parameterize heat transport in the low-$\Ron$ regime relevant to icy moon oceans. The scheme captures the meridional heat transport associated with slantwise convection, which is absent in traditional convection schemes. This transport modulates the surface heat flux by up to a factor of 2 relative to the imposed bottom heat flux. Such redistribution affects the ice-ocean heat budget and may influence ice shell topography, one of the key observables of icy moons.

The scheme performs best inside the tangent cylinder, where it accurately reproduces the meridional heat transport from convection-resolving simulations, and performs reasonably well outside the tangent cylinder where it captures the sign and general magnitude of the meridional heat transport. Our sensitivity tests show that heat transport outside the tangent cylinder remains sensitive to the resolved dynamics, while heat transport inside the tangent cylinder is reproduced robustly across a wide range of resolutions and viscosities. The reduced accuracy outside the tangent cylinder likely reflects the distinct dynamical regime dominated by equatorial-roll-like motions. Future work should therefore focus on understanding these dynamics and developing improved parameterizations for them.

Although the slantwise convection scheme performs well in representing the bulk vertical temperature gradient in the ocean interior, it does not fully reproduce the temperature field in the convection-resolving simulations. The likely explanation is that the scheme does not represent the boundary layers, which set the boundary conditions for temperature and heat transport in the interior. However, the viscous boundary layers along a smooth surface in the convection-resolving simulations are themselves unrealistic, as even the CR simulations cannot adequately resolve the boundary layer dynamics, and the boundaries of real icy moon oceans are likely rough rather than smooth. Future studies should therefore investigate appropriate boundary layer parameterizations for icy moon oceans and their influence on the temperature structure and heat transport.

In addition, the scheme does not account for momentum transport by slantwise convection, which may lead to nonsynchronous rotation of ice shells \citep{ashkenazy2023non,kang2024nonsynchronous}. Extending the scheme to include momentum redistribution by slantwise convection is therefore an important direction for future work.

In the benchmark tests, density is controlled solely by temperature. In realistic icy moon oceans, salinity can also play an important role \citep{zeng2021ocean,kang2022does}. The scheme can be directly extended to include salinity under a linear equation of state. However, the equation of state for seawater under icy moon conditions can be strongly nonlinear, and the impact of this nonlinearity on convective transport remains to be explored. The current formulation assumes that convection spans the full ocean depth. Previous studies suggest that for Enceladus, a stably stratified layer may exist above a convective layer when salinity is low \citep{zeng2021ocean,zeng2024effect,ames2025ocean}. In such cases, the depth scale for slantwise convection $H_z$ should be reduced to the depth of the convective layer, which can be estimated from the bottom heat flux and background vertical diffusivity (cf. Eq.~1 in \cite{zeng2021ocean}).

This study isolates the effect of slantwise convection and neglects baroclinic eddies. However, previous work has shown that baroclinic eddies can dominate heat transport when the ice shell is not flat \citep{kang2023modulation,zeng2024effect}. Future work should therefore examine the interaction between slantwise convection and baroclinic eddies. Because slantwise convection is not resolved in global simulations, its effects must be parameterized. The scheme developed here can be combined with explicitly resolved baroclinic eddies in global models, providing a framework for representing both processes and distinguishing their contributions to ocean heat transport.

Beyond icy moons, this scheme can also be applied to other oceans in the rapidly rotating regime, i.e., $\Ron \equiv B^{1/2} (2\Omega)^{-3/2}H^{-1} \ll 1$. This regime can be realized when the ocean depth $H$ is large, and may therefore be relevant to deep ocean worlds on exoplanets \citep{kuchner2003volatile,leger2004new,sotin2007mass,kite2018habitability,lai2022thermocline}.

\section*{Acknowledgements}

We thank Wanying Kang, John Marshall, Kaushal Gianchandani, and Hing Ong for helpful discussions and comments. This work was completed using resources provided by the University of Chicago Research Computing Center.

\section*{Data Availability}

The MITgcm model with the slantwise convection parameterization implemented is available at \url{https://github.com/yaoxuanzeng/MITgcm/releases/tag/slantwise_convection_scheme_v1.0}. The simulation data analyzed in this study are available on Zenodo:
\url{https://doi.org/10.5281/zenodo.20773368}.



\bibliographystyle{mnras}
\bibliography{reference} % if your bibtex file is called example.bib

%\begin{thebibliography}{99}
%\end{thebibliography}

%%%%%%%%%%%%%%%%%%%%%%%%%%%%%%%%%%%%%%%%%%%%%%%%%%

%%%%%%%%%%%%%%%%% APPENDICES %%%%%%%%%%%%%%%%%%%%%

\newpage

\appendix

\section{An alternative choice of $H_z$ outside the tangent cylinder}\label{app:Hz}

\begin{figure}
    \centering
    \includegraphics[width=0.96\textwidth]{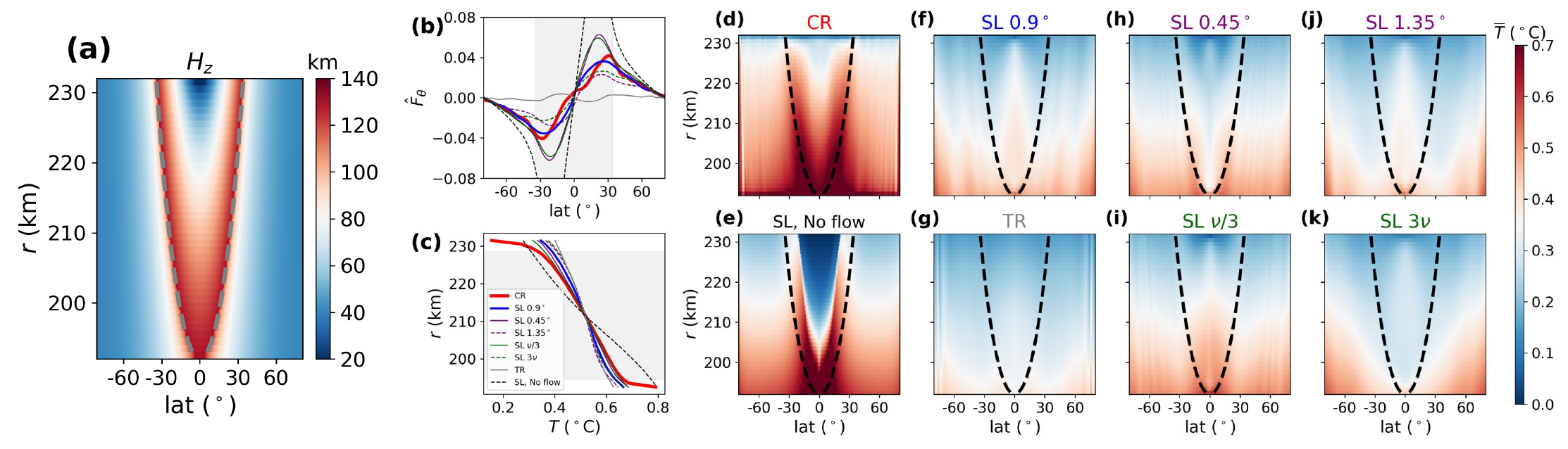}
    \caption{\textbf{Heat transport and temperature fields in the sensitivity tests with $\Ron=5.6 \times 10^{-3}$ and varying $H_z$ outside the tangent cylinder.} Panels (b)-(k) are the same as Fig.~\ref{fig:sensitivity:heat}, but with varying $H_z$ outside the tangent cylinder (see panel a).}
    \label{fig:sensitivity:heat_oldHz}
\end{figure}

The appropriate length scale $H_z$ outside the tangent cylinder is not well constrained, as the dynamics there are dominated by equatorial rolls rather than slantwise convection. In the default setup, we prescribe a constant $H_z$ outside the tangent cylinder. An obvious alternative choice is to define $H_z$ as the distance between the ice shell and the equatorial plane, which yields an $H_z$ profile that decreases to zero toward the equator (Fig.~\ref{fig:sensitivity:heat_oldHz}a).

To assess the sensitivity to this choice, we rerun $\Ron=5.6\times10^{-3}$ simulations using the alternative $H_z$ profile while keeping all other parameters unchanged. The resulting meridional heat transport and temperature structure remain broadly unchanged (compare Fig.~\ref{fig:sensitivity:heat} and Fig.~\ref{fig:sensitivity:heat_oldHz}). The alternative formulation produces slightly better agreement with the convection-resolving simulation in the meridional heat transport outside the tangent cylinder. However, it yields a less realistic temperature structure in some simulations, with a temperature maximum outside the tangent cylinder rather than the temperature minimum seen in the convection-resolving simulation.

\section{Rayleigh number and plume scale}\label{app:Ra_Ek}

\begin{table}
\caption{\textbf{Plume scales and diagnosed Rayleigh numbers for the convection-resolving simulations.} The average ocean radius, $\overline{r} = (r_{\mathrm{core}} + r_{\mathrm{surf}})/2$, is used to convert the estimated plume size into an equivalent latitudinal scale.}
\centering
\begin{tabular}{c|ccc|c}
\toprule
$\Ron$ & $\mathrm{Ra}= g \alpha_T \Delta T H^3 \nu^{-1} \kappa^{-1}$ & $\mathrm{Ra_{T,RR}}=0.4 \mathrm{Ek}^{-8/5}$ & $\mathrm{Ra_{T,NR}}=100 \mathrm{Ek}^{-12/7}$ & $\Delta \theta_{plume} \sim 5B^{1/4} (2\Omega)^{-3/4}H^{1/2}/\overline{r}$ \\
\midrule
$5.6 \times 10^{-2}$ & $2.5 \times 10^7$ & $6.0 \times 10^{4}$ & $3.5 \times 10^7$ & 12.8$^\circ$  \\
$1.8 \times 10^{-2}$ & $4.1 \times 10^7$ & $2.1 \times 10^{5}$ & $1.3 \times 10^8$ & 7.2$^\circ$ \\
$5.6 \times 10^{-3}$ & $5.7 \times 10^7$ & $7.0 \times 10^{5}$ & $4.9 \times 10^8$ & 4.1$^\circ$ \\
$1.8 \times 10^{-3}$ & $9.9 \times 10^7$ & $2.4 \times 10^{6}$ & $1.8 \times 10^9$ & 2.3$^\circ$ \\
\bottomrule
\end{tabular}
\label{tab:Ra_Ek}
\end{table}

We estimate the convective plume scale in the simulations as $l_{\mathrm{plume}}\sim B^{1/4} (2\Omega)^{-3/4}H^{1/2}$ \citep{fernando1989turbulent,jones1993convection}. The actual plume size is larger than this scaling by a prefactor, reported to be approximately 20 in \cite{bire2022exploring} and 5 in \cite{zeng2026slantwise}. Taking this prefactor into account, to ensure that convection plumes are adequately resolved in the CR simulations, we choose horizontal resolutions smaller than $l_{\mathrm{plume}}/10$ (Table~\ref{tab:Ra_Ek}).

Using the posteriori diagnosed horizontally-averaged vertical temperature contrast $\Delta T$, we calculate the Rayleigh number for the convection-resolving simulations (Table~\ref{tab:Ra_Ek}). For comparison, we also compute the transitional Rayleigh numbers $\mathrm{Ra_{T,RR}} = 0.4 \mathrm{Ek}^{-8/5}$ and $\mathrm{Ra_{T,NR}} = 100 \mathrm{Ek}^{-12/7}$, which represent the transitions from the rapidly rotating regime to the transitional regime and from the transitional regime to the non-rotating regime, respectively \citep{gastine2016scaling}.

For the simulation with $\Ron = 5.6 \times 10^{-2}$, we find $\mathrm{Ra} \sim \mathrm{Ra_{T,NR}}$, indicating that this case lies near the non-rotating regime. In this regime, convective flows are dominated more by buoyancy than by rotation, consistent with the absence of rotation-aligned structures and the tendency for plumes to align with gravity. 

For the simulations with $\Ron = 5.6 \times 10^{-3}$ and $1.8 \times 10^{-3}$, we find $\mathrm{Ra_{T,RR}} < \mathrm{Ra} < \mathrm{Ra_{T,NR}}$, with $\mathrm{Ra}$ approximately an order of magnitude away from both transition values, indicating that these cases lie within the transitional regime. In this transitional regime, the interior flow remains strongly influenced by rotation, while the boundary layers are not rotationally constrained. This is consistent with the presence of slantwise plumes aligned with the rotation axis in the interior. Icy moon oceans are expected to lie within this transitional regime. The good agreement in meridional heat transport for these low-$\Ron$ simulations therefore supports the applicability of the slantwise convection scheme developed in this study to realistic icy moon conditions.

%%%%%%%%%%%%%%%%%%%%%%%%%%%%%%%%%%%%%%%%%%%%%%%%%%

% Don't change these lines
\bsp	% typesetting comment
\label{lastpage}
\end{document}